\documentclass[preprint]{jpsj2}

\def\runauthor{
Kenji \textsc{Yonemitsu}
}

\title{
Phase Transition in a One-Dimensional Extended Peierls-Hubbard Model with a Pulse of Oscillating Electric Field: II. Linear Behavior in Neutral-to-Ionic Transition
}

\author{
Kenji \textsc{Yonemitsu}$^{1,2}$
\thanks{E-mail address: kxy@ims.ac.jp}
}

\inst{
$^1$Institute for Molecular Science, Okazaki 444-8585 \\
$^2$Graduate University for Advanced Studies, Okazaki 444-8585
}

\recdate{\today}

\abst{
Dynamics of charge density and lattice displacements after the neutral phase is photoexcited is studied by solving the time-dependent Schr\"odinger equation for a one-dimensional extended Peierls-Hubbard model with alternating potentials. In contrast to the ionic-to-neutral transition studied previously, the neutral-to-ionic transition proceeds in an uncooperative manner as far as the one-dimensional system is concerned. The final ionicity is a linear function of the increment of the total energy. After the electric field is turned off, the electronic state does not significantly change, roughly keeping the ionicity, even if the transition is not completed, because the ionic domains never proliferate. As a consequence, an electric field with frequency just at the linear absorption peak causes the neutral-to-ionic transition the most efficiently. These findings are consistent with the recent experiments on the mixed-stack organic charge-transfer complex, TTF-CA. We artificially modify or remove the electron-lattice coupling to discuss the origin of such differences between the two transitions. 
}

\kword{
photoinduced phase transition, neutral-ionic transition, TTF-CA, charge-transfer complex, time-dependent Schr\"odinger equation, dynamics
}

\begin{document}
\sloppy
\maketitle

\section{Introduction}

In the previous paper \cite{yonemitsu04a}, we treat the charge-lattice-coupled dynamics from the ionic to neutral phases. It corresponds to the transition from the low-temperature phase with spontaneously broken, inversion symmetry accompanied by the long-range order of polarizations to the high-temperature phase with the restored symmetry, although the transition is not triggered by modulation of the temperature. In general, such a transition is supposed to be more easily achieved than the reverse transition. Even in the TTF-CA complex, which we presently focus on, it was regarded as difficult in the initial stage of the experimental studies to realize a transition from the neutral phase to the ionic one. Therefore, there would be some differences between the ionic-to-neutral and neutral-to-ionic transitions. Now both transitions are observed \cite{koshihara99}. The x-ray diffraction clearly demonstrates that the transition from the neutral phase to the ionic phase with three-dimensionally ordered polarizations is induced by intramolecular excitations \cite{collet03}.

In addition to the difference in the inversion symmetry, the electronic states in the TTF-CA complex have different degrees of electron correlation. The neutral state is a band insulator, while the ionic state is basically a Mott insulator, although the latter state is also nonmagnetic due to the dimerization of donor and acceptor molecules. The Mott insulator at half filling is of course a consequence of the electron correlation that tends to avoid double occupancy of molecular orbitals. In this context, it would be highly desirable to theoretically study whether any difference is expected in these transitions. In fact, qualitative differences are observed in the calculated dynamics of charge density and that of lattice displacements in the one-dimensional model system.

In recent reflectance studies after a pulse of intra-chain charge transfer excitation, although the ionic-to-neutral transition shows a threshold density of absorbed photons \cite{iwai02}, the dynamics from the neutral phase toward the ionic one initially shows that the conversion ratio is an almost linear function of the density of absorbed photons \cite{okamoto_unpublished}. In the latter dynamics, the conversion ratio decreases with time, and the complex finally returns to the neutral phase with additional thermal fluctuations. Thus, a transition to the long-living ionic phase is not achieved by intra-chain charge transfer excitations. In the most recent studies, it is not realized by intramolecular excitations, either \cite{okamoto_unpublished2}. The reason for the discrepancy between this result and that of ref.~\citen{collet03} is not clear. 

It is true that the difference in the absorption depths between intra-chain charge transfer excitations and intramolecular excitations was regarded as important. Photons with intramolecular excitation energies are more weakly absorbed and penetrate the sample more deeply, which may help the transition to the ionic phase. But it is confirmed in the experiments in ref.~\citen{okamoto_unpublished2} that the thickness of the sample is much larger than both of these absorption depths. The reason for the discrepancy is unknown at the moment. In any case, photons with intra-chain charge transfer excitations have not so far caused any transition to the long-living ionic phase in the TTF-CA complex. We will then compare the numerical results corresponding to intra-chain charge transfer excitations with this experimental result. 

The x-ray diffraction shows the averaged bulk property \cite{collet03}, while the reflectance generally has a better space and time resolution \cite{okamoto_unpublished,okamoto_unpublished2} but is more sensitive to the surface property. It is noted here that, although the transition mechanism is very different from the present material, the spin-crossover complex [Fe(2pic)$_3$]Cl$_2$EtOH may show photoinduced properties \cite{collet_spin_crossover} that are very different from thermally induced ones \cite{chernyshov_spin_crossover} often characterized by the two-step transition, possibly depending on the space and time scales as discussed theoretically \cite{luty_yonemitsu_spin_crossover}. Searching a possibility or mechanism of scale-dependent properties is a general subject and left for future studies. 

In this paper, we show the characters of the neutral-to-ionic transition that contrast with those of the ionic-to-neutral transition presented in the previous paper \cite{yonemitsu04a}. The ionicity after the transient period is a linear function of the increment of the total energy. After the electric field is turned off, the electronic state does not significantly change, roughly keeping the ionicity, even if the transition is not completed. This implies that the cooperativity observed in the ionic-to-neutral transition does not appear in the neutral-to-ionic transition of the isolated one-dimensional model system. In the real material, energy dissipation would easily convert it back into the neutral phase. 

At present, different types of photoinduced phase transitions are experimentally known. In some cases the symmetry of the electronic state is unchanged, e.g., during the transition from the Mott insulator phase to the metallic phase in halogen-bridged nickel complexes \cite{okamoto_Mott_metal}, while in other cases the symmetry is restored, e.g., after the transition from the charge-density-wave phase to the Mott insulator phase in halogen-bridged palladium complexes \cite{okamoto_CDW_Mott}. In halogen-bridged {\it binuclear} platinum complexes, the charge-density-wave phase before and the charge-polarization phase after the transition have spontaneously broken, different symmetries \cite{matsuzaki_CDW_CP}, for which the electronic mechanism is theoretically clarified \cite{yonemitsu03}. The time scales of the transitions mentioned above are very different. Their dynamical characters are not fully studied yet, but they will be clarified in future. The variety of photoinduced phase transitions is not simply caused by different symmetries. The lattice degrees of freedom would be important in determining the time scale. They are not coupled with the system in the rapid transition from the Mott insulator phase to the metallic phase, while they are strongly coupled in the others. The electron correlation is crucial in the transitions from or to the Mott insulator phase, where spin fluctuations may be deeply involved with the transition dynamics. 

To systematically develop theories for these photoinduced phase transitions, the TTF-CA complex will continue to be important as a typical example because its dynamical characters are the most extensively studied, which allows comparisons with theoretical calculations, and because all of the symmetry, the involvement of lattice displacements, and the degree of electron correlation are largely changed through the transition. Although the present calculations are limited to isolated one-dimensional model systems, i.e., without dissipation or inter-chain couplings, they will contribute to future studies to seek the way to control the photoinduced phase transition dynamics by choosing the symmetry, coupling or electron correlation.

\section{Extended Peierls-Hubbard Model with Alternating Potentials}\label{model}

We use a one-dimensional extended Peierls-Hubbard model with alternating potentials at half filling, as in paper I,
\begin{equation}
H =  H_\mathrm{el} + H_\mathrm{lat} \;,\label{g-ham} 
\end{equation}
with
\begin{align}
H_\mathrm{el} = & 
-t_0 \sum _{\sigma,l=1}^{N}
   \left( c^{\dagger }_{l+1,\sigma}c_{l,\sigma}
     + \mathrm{h.c.} \right)  \nonumber \\ 
& +\sum _{l=1}^{N} 
\left[ U n_{l,\uparrow} n_{l,\downarrow} 
+ (-1)^{l} \frac{d}{2} n_{l} \right] \nonumber \\ 
& +\sum _{l:odd}^{N} 
\bar{V}_{l} (n_{l}-2) n_{l+1}
+ \sum _{l:even}^{N} 
\bar{V}_{l} n_{l} (n_{l+1}-2) \;, \label{e-ham} \\
H_\mathrm{lat} =& \sum _{l=1}^{N}
\left[ \frac{k_1}{2}y_{l}^{2}
+\frac{k_2}{4}y_{l}^{4}
+\frac{1}{2}m_{l}\dot{u}_{l}^{2} \right] \;,\label{l-ham} 
\end{align}
where, $ c^{\dagger }_{l,\sigma} $  ($ c_{l,\sigma} $) is the creation (annihilation) operator of a $\pi$-electron with spin $\sigma$ at site $l$, $ n_{l,\sigma} = c^{\dagger}_{l,\sigma} c_{l,\sigma} $, $ n_{l} = n_{l,\uparrow} + n_{l,\downarrow} $, $ u_{l} $ is the dimensionless lattice displacement of the $l$th molecule along the chain from its equidistant position, and $ y_{l} = u_{l+1} - u_{l} $.
The distance between the $l$th and $(l+1)$th molecules is then given by $r_l=r_0 (1+u_{l+1}-u_l)$, where $r_0$ is the averaged distance between the neighboring molecules along the chain. 
The other notations are also the same as in paper I. 
We numerically solve the time-dependent Schr\"odinger equation with the help of the unrestricted Hartree-Fock approximation for the electronic part, and the classical equation of motion for the lattice part, as described in paper I. 
Note that random numbers are added to the initial $ y_{l} $ and $ \dot{u}_l $  values according to the Boltzmann distribution at a fictitious temperature $\it{T}$. 

Photoexcitations are introduced by modifying the transfer integral with the Peierls phase, as explained in paper I. 
The pulse of electric field is given by 
\begin{equation}
E(t) = E_{\rm ext} \sin \omega_{\rm ext} t
\;,
\end{equation}
with amplitude $ E_{\rm ext} $ and frequency $ \omega_{\rm ext} $ for $ 0 < t < N_{\rm ext} T_{\rm ext} $ with integer $ N_{\rm ext} $. $ E(t) $ is zero otherwise. The period is denoted by $ T_{\rm ext} $, which is given by $ T_{\rm ext} = 2 \pi / \omega_{\rm ext} $.
Note that this photoexcitation is different from ref.~\citen{miyashita03}.

\section{Results and Discussions}\label{results}

We use $N$=100, $ t_0 $=0.17eV, $ U $=1.528eV, $ V $=0.600eV, $ d $=2.716eV, $ \beta_2 $=8.54eV, $ k_1 $=4.86eV, $ k_2 $=3400eV, and the bare phonon energy $ \omega_{\rm opt} \equiv (1/r_0)(2 k_1/m_r)^{1/2} $=0.0192eV, and impose the periodic boundary condition.
Here $m_r$ is the reduced mass for the donor and acceptor molecules. 
Now only the value of $ V $ is different from and slightly smaller than that in paper I. With these parameters, the neutral phase is now stable and the dimerized ionic phase is metastable.
The ionicity is defined as $ \rho = 1 + (1/N)  \sum_{l=1}^{N} (-1)^{l} \langle n_{l} \rangle $.
The staggered lattice displacement is defined as $ y_{st} = (1/N)  \sum_{l=1}^{N} (-1)^l y_l $.

\subsection{Linear absorption}\label{subsec:linear_absorption}

\begin{figure}
\includegraphics[height=6cm]{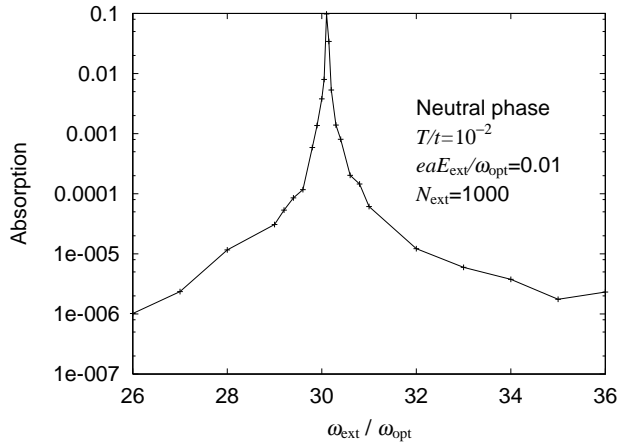}
\caption{ Linear absorption, as a function of $ \omega_{\rm ext} / \omega_{\rm opt} $, in the neutral phase with lattice fluctuations at $ T/t $=10$^{-2}$. The applied electric field is very weak, $ eaE_{\rm ext}/\omega_{\rm opt} $=0.01, and lasts long, $ N_{\rm ext} $=1000.}
\label{fig:linear_absorption}
\end{figure}

The linear absorption spectrum is calculated by applying a very weak electric field for a long time and by observing the increment of the total energy (Fig.~\ref{fig:linear_absorption}). Since $ \omega_{\rm opt} $=0.0192eV is used here, the position of the absorption peak is $ \omega_{\rm ext} \simeq 30 \omega_{\rm opt} \simeq $0.58eV. Compared with the spectrum in the ionic phase, that in the neutral phase is much less sensitive to lattice fluctuations because the electron-lattice coupling is considered only in the attraction between an electron and a hole on the neighboring acceptor and donor sites, whose densities are both low in the neutral phase. Lattice fluctuations used here correspond to $ T/t $=10$^{-2}$, i.e., $ T/\omega_{\rm opt} \simeq $0.089.

\subsection{Nonlinear absorption and linear behavior}\label{subsec:nonlinear_absorption}

\begin{figure}
\includegraphics[height=12cm]{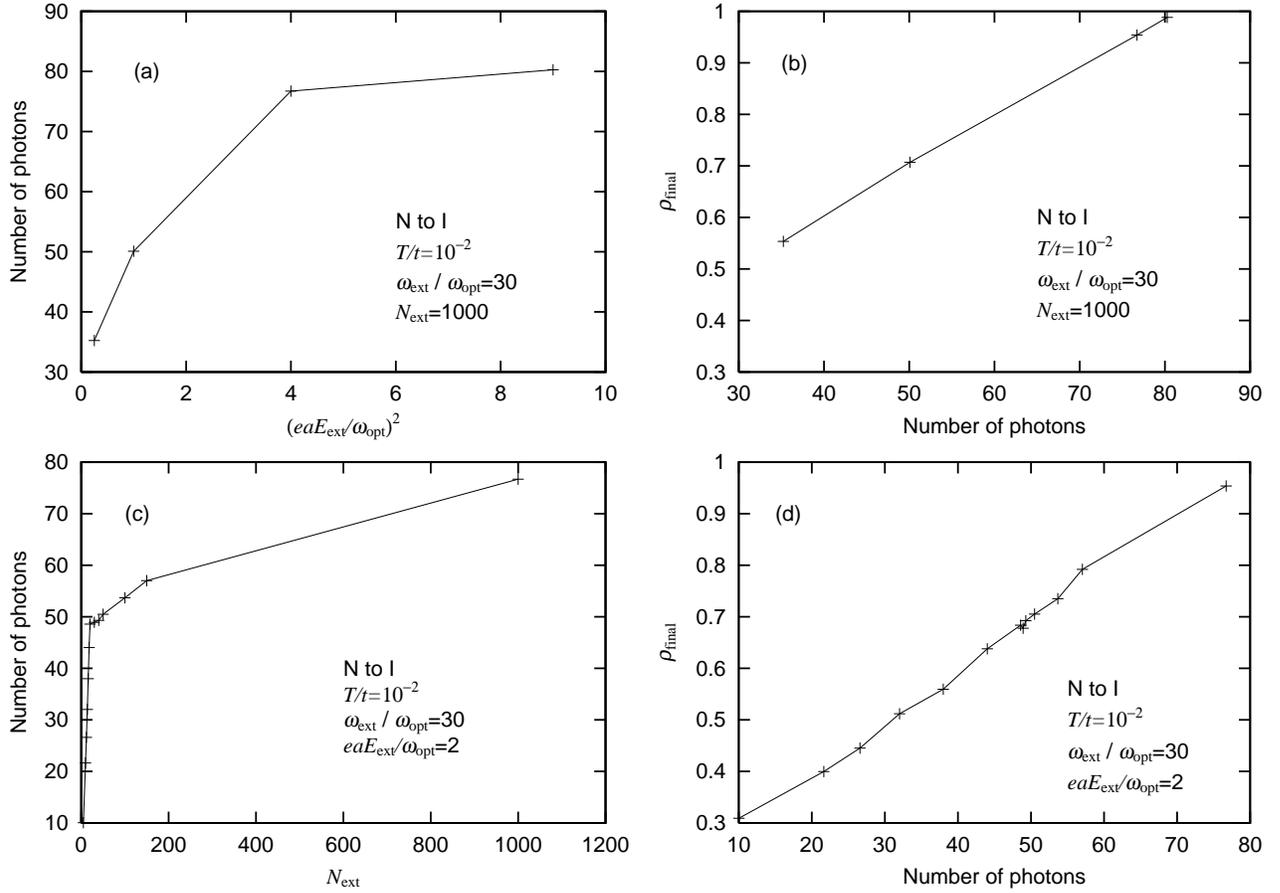}
\caption{(a) Number of absorbed photons as a function of the field strength $ ( eaE_{\rm ext}/\omega_{\rm opt} )^2 $, and (b) the corresponding final ionicity, for $ N_{\rm ext} $=1000. (c) Number of absorbed photons as a function of the pulse duration $ N_{\rm ext} $, and (d) the corresponding final ionicity, for $ eaE_{\rm ext}/\omega_{\rm opt} $=2. The electric field of frequency $ \omega_{\rm ext} / \omega_{\rm opt} $=30 near the linear-absorption peak is applied to the neutral phase at $ T/t $=10$^{-2}$.}
\label{fig:pulse_dependence_on_resonance}
\end{figure}

By setting the frequency of the electric field $ \omega_{\rm ext} $ near the linear-absorption peak, we calculate the dependence of the number of absorbed photons and that of the final ionicity on the field strength $ ( eaE_{\rm ext} )^2 $ and the pulse duration $ N_{\rm ext} T_{\rm ext} $, which are shown in Fig.~\ref{fig:pulse_dependence_on_resonance}. Because the field is almost on resonance, the number of absorbed photons is a rapidly increasing function of the strength or of the duration when it is small.
With increasing the strength or the duration, the number of absorbed photons roughly approaches the number of sites and tends to be saturated.
In spite of the highly nonlinear dependence on the strength or the duration, the final ionicity is almost linearly related with the number of absorbed photons.
In other words, the final state is determined merely by how much photons are absorbed to increase the ionicity. The cooperative character is missing here. This linear behavior is in contrast to the threshold behavior in the ionic-to-neutral transition. 

\begin{figure}
\includegraphics[height=12cm]{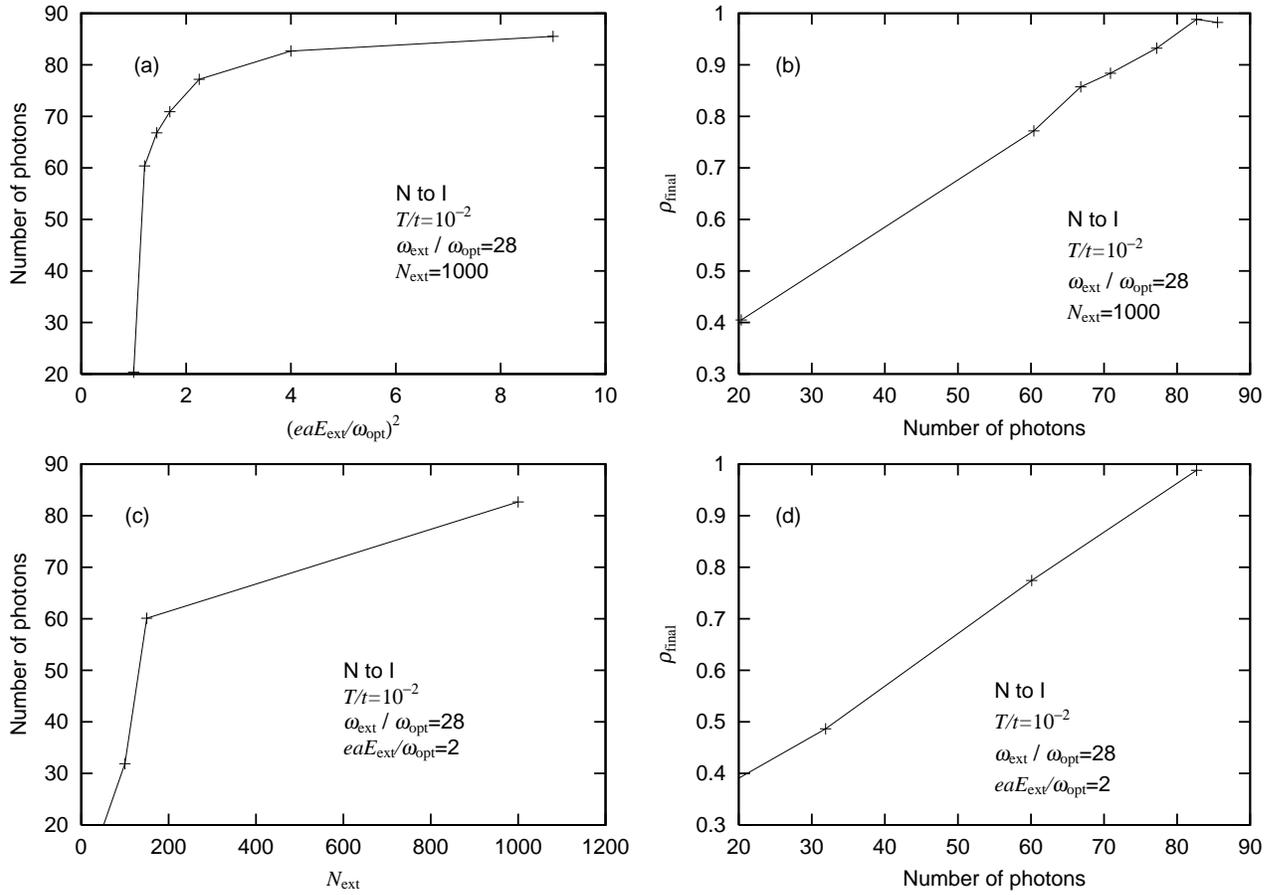}
\caption{(a) Number of absorbed photons as a function of the field strength $ ( eaE_{\rm ext}/\omega_{\rm opt} )^2 $, and (b) the corresponding final ionicity, for $ N_{\rm ext} $=1000. (c) Number of absorbed photons as a function of the pulse duration $ N_{\rm ext} $, and (d) the corresponding final ionicity, for $ eaE_{\rm ext}/\omega_{\rm opt} $=2. The electric field of frequency $ \omega_{\rm ext} / \omega_{\rm opt} $=28 below the linear-absorption peak is applied to the neutral phase at $ T/t $=10$^{-2}$.}
\label{fig:pulse_dependence_off_resonance}
\end{figure}

To see how the behavior could be modified when the field is off resonance, we then use a frequency $ \omega_{\rm ext} $ below the linear-absorption peak in Fig.~\ref{fig:pulse_dependence_off_resonance}. If the field is too weak or too short, photons are hardly absorbed due to the off-resonant excitation. Except for this fact, the behavior is very similar to that in the on-resonant case. Above some finite value of strength or duration, the number of absorbed photons rapidly increases. Then, it roughly approaches the number of sites and tends to be saturated. The final ionicity is again linearly related with the number of absorbed photons. Note that all the relations between the final ionicity and the number of absorbed photons, Figs.~\ref{fig:pulse_dependence_on_resonance}(b), \ref{fig:pulse_dependence_on_resonance}(d), \ref{fig:pulse_dependence_off_resonance}(b) and \ref{fig:pulse_dependence_off_resonance}(d), with different strengths or durations for on- and off-resonant excitations, show almost a common linear coefficient ($ \rho_{\rm final} $=0.4, 0.6, 0.8 and 1 roughly when 20, 40, 60 and 80 photons are absorbed in the 100-site chain).

When Fig.~\ref{fig:pulse_dependence_off_resonance}(c) is compared with Fig.~\ref{fig:pulse_dependence_on_resonance}(c), we see that a much longer pulse is needed for the 100-site chain to absorb 50 photons if it is off resonance. It means that the on-resonant pulse is much more efficient with a given strength and a given duration than the off-resonant pulse. This fact is also in contrast to that in the ionic-to-neutral transition where the off-resonant pulse is more efficient than the on-resonant condition. The linear dependence of the outcome on the absorbed energy suggests that the charge transfer takes place uncooperatively. Then, the most efficient pulse is that which is the most strongly absorbed when the neutral phase is photoexcited. The uncooperative character will be clarified in the following by the spatial and temporal development of the ionicity. 

\begin{figure}
\includegraphics[height=12cm]{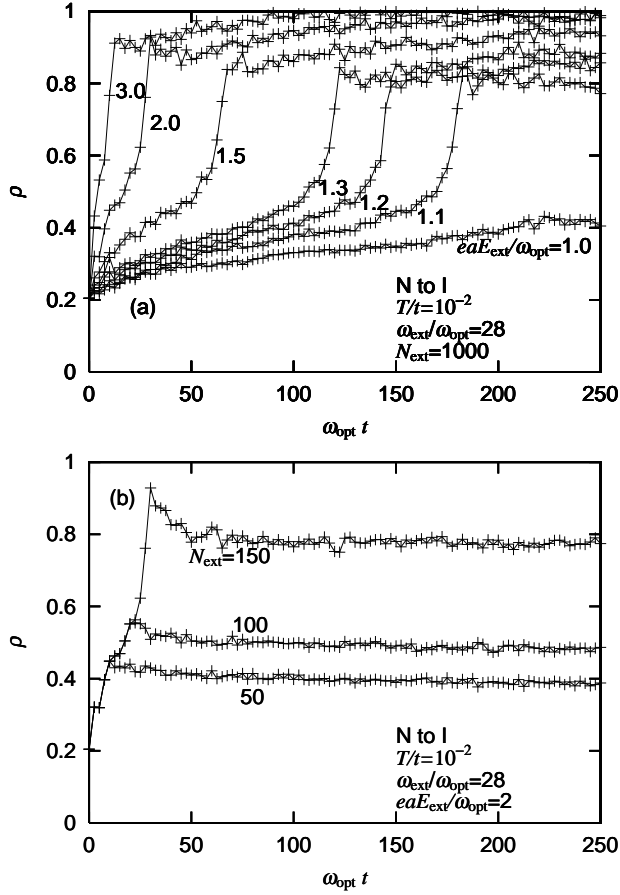}
\caption{Time dependence of the ionicity, (a) with different field strengths $ ( eaE_{\rm ext}/\omega_{\rm opt} )^2 $ for $ N_{\rm ext} $=1000, and (b) with different pulse durations $ N_{\rm ext} $ for $ eaE_{\rm ext}/\omega_{\rm opt} $=2. The electric field of frequency $ \omega_{\rm ext} / \omega_{\rm opt} $=28 below the linear-absorption peak is applied [(a) for $ 0 \leq \omega_{\rm opt} t \leq N_{\rm ext} \omega_{\rm opt} T_{\rm ext} $=224, and (b) for $ 0 \leq \omega_{\rm opt} t \leq N_{\rm ext} \omega_{\rm opt} T_{\rm ext} $=11.2, 22.4 and 33.7 with $ N_{\rm ext} $=50, 100 and 150, respectively] to the neutral phase at $ T/t $=10$^{-2}$. The pulse duration required for the transition changes continuously with the field strength [(a)]. The ionicity is almost unchanged after the field is turned off [(b)].}
\label{fig:time_dependence_off_resonance}
\end{figure}

The evolution of the ionicity is shown in Fig.~\ref{fig:time_dependence_off_resonance}(a) with different field strengths $ ( eaE_{\rm ext}/\omega_{\rm opt} )^2 $ and in Fig.~\ref{fig:time_dependence_off_resonance}(b) with different pulse durations $ N_{\rm ext} $. In Fig.~\ref{fig:time_dependence_off_resonance}(a), the oscillating electric field is applied for $ 0 \leq \omega_{\rm opt} t \leq 224 $, i.e., almost all the time except the last 10\%. With weak fields, the ionicity increases a little bit, but the system remains neutral. Above some finite strength for a given duration, the rapid and large increase of the ionicity is observed before the field is turned off, and the system is converted into an ionic phase. As the field is strengthened, the rapid increase takes place earlier. Strong fields increase the ionicity rapidly from the beginning. Although the increase itself is rapid, the overall behavior is changed continuously with increasing strength. In Fig.~\ref{fig:time_dependence_off_resonance}(b), the field is applied for $ 0 \leq \omega_{\rm opt} t \leq 11.2$, 22.4 and 33.7 with $ N_{\rm ext} $=50, 100 and 150, respectively. When the field is turned off during the rapid increase of the ionicity, the ionicity decreases a little bit immediately after that, but it roughly keeps a value which the system had when the field is turned off. Now the linear behavior shown in Figs.~\ref{fig:pulse_dependence_on_resonance} and \ref{fig:pulse_dependence_off_resonance} is found to be a consequence of the very fact that the ionicity remains an intermediate value if the field is turned off before the transition is completed. 

\begin{figure}
\includegraphics[height=6cm]{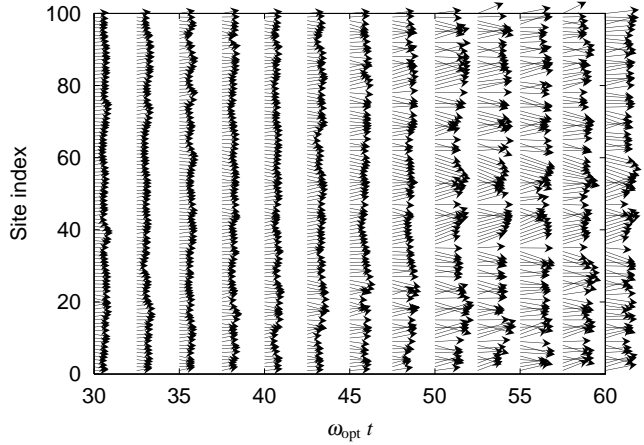}
\caption{Correlation between the staggered lattice displacement $ y_{st \; l} $ (the vertical component of the bar) and the ionicity $ \rho_l $ (the horizontal component of the bar), as a function of the site index $ l $ and the elapsing time $ t $ multiplied by $ \omega_{\rm opt} $. The electric field with $ eaE_{\rm ext}/\omega_{\rm opt} $=2, $ N_{\rm ext} $=1000 and of frequency $ \omega_{\rm ext} / \omega_{\rm opt} $=28 below the linear-absorption peak is applied (for $ 0 \leq \omega_{\rm opt} t \leq N_{\rm ext} \omega_{\rm opt} T_{\rm ext} $=224) to the neutral phase at $ T/t $=10$^{-3}$. The final state is ionic with short-range-ordered lattice alternation.}
\label{fig:space_time_dependence}
\end{figure}

The space and time evolution of the ionicity and the staggered lattice displacement is shown in Fig.~\ref{fig:space_time_dependence}. The horizontal component of each bar represents the local ionicity $ \rho_l $ defined as $ \rho_l = 1 + (-1)^l ( - \langle n_{l-1} \rangle + 2 \langle n_l \rangle - \langle n_{l+1} \rangle )/4 $. The vertical component gives the local staggered lattice displacement $ y_{st \; l} $ defined as $ y_{st \; l} = (-1)^l ( - y_{l-1} + 2 y_l - y_{l+1} )/4 $. The bars are shown on all sites and selected times. Now the oscillating electric field is applied during the time span shown in the figure. In this time span, the ionicity changes the most rapidly. Nevertheless the local ionicity increases much more gradually than in the ionic-to-neutral transition, and more significantly, their increase is rather uniform. As a consequence, the boundary between the ionic and neutral domains is not clearly visible. This behavior is in clear contrast to that in the ionic-to-neutral transition, where the boundary is clearly visible, and the transition spontaneously proceeds once triggered. This fact enables the ionicity to remain an intermediate value if the field is turned off before the transition is completed. Later, we show the effect of spin fluctuations, which produce some partially ionic domains and make the domain boundaries visible, but the transition never proceeds spontaneously even in such a condition. Therefore, the difference between the neutral-to-ionic and ionic-to-neutral transitions is not due to the neglect of spin fluctuations in the present calculations. As the ionicity increases, the positions of the molecules deviate from the equidistant ones. Then the lattice displacement is locally staggered, but not globally. Then, ionic domains with different polarizations coexist. These domains continue to compete with each other and none of them finally predominates. 

\subsection{Effect of spin fluctuations}\label{subsec:spin_fluctuations}

So far, random numbers have been added only to the initial lattice variables. If we start to solve the time-dependent Schr\"odinger equation from the neutral phase, then the up-spin and down-spin one-body wave functions remain degenerate all the time because the electric field does not lift the degeneracy. Thus there is no chance for a magnetic moment to appear. The states to be reached from the neutral phase have so far been different from the ionic state, where the magnetic moments are staggered due to the mean-field approximation. This difference is partially overcome by introducing randomness not only to the lattice variables but also to the spin densities. We then add in this subsection very small random numbers to the initial spin densities so that the increment of the total energy is comparable with that due to the random numbers added to the lattice variables. Then, the system spontaneously recovers finite staggered magnetization in the ionic phase, although it is an artifact of the approximation. The inclusion of spin fluctuations promotes nucleation of ionic domains and accelerates the transition to the ionic phase

\begin{figure}
\includegraphics[height=12cm]{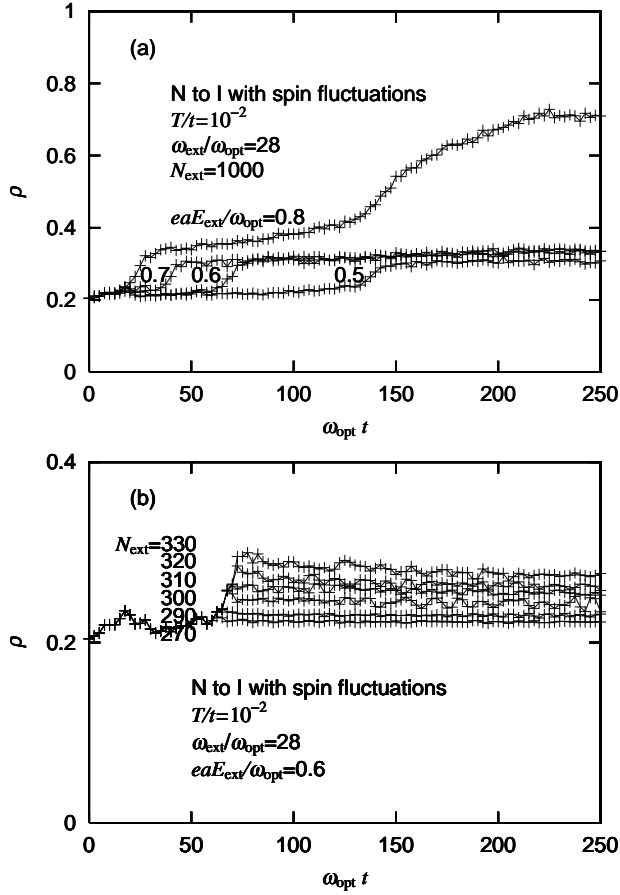}
\caption{Time dependence of the ionicity, (a) with different field strengths $ ( eaE_{\rm ext}/\omega_{\rm opt} )^2 $ for $ N_{\rm ext} $=1000, and (b) with different pulse durations $ N_{\rm ext} $ for $ eaE_{\rm ext}/\omega_{\rm opt} $=0.6. The electric field of frequency $ \omega_{\rm ext} / \omega_{\rm opt} $=28 is applied [(a) for $ 0 \leq \omega_{\rm opt} t \leq N_{\rm ext} \omega_{\rm opt} T_{\rm ext} $=224, and (b) for $ 0 \leq \omega_{\rm opt} t \leq N_{\rm ext} \omega_{\rm opt} T_{\rm ext} $=60.6, 65.1, 67.3, 69.6, 71.8 and 74.1 with $ N_{\rm ext} $=270, 290, 300, 310, 320 and 330, respectively] to the neutral phase with spin fluctuations at $ T/t $=10$^{-2}$.}
\label{fig:time_dependence_spin_fluctuation}
\end{figure}

After the inclusion of both lattice and spin fluctuations in the initial state, the ionicity evolves as shown in Fig.~\ref{fig:time_dependence_spin_fluctuation}(a) for different field strengths $ ( eaE_{\rm ext}/\omega_{\rm opt} )^2 $ and in Fig.~\ref{fig:time_dependence_spin_fluctuation}(b) for different pulse durations $ N_{\rm ext} $. Note that the oscillating electric field is applied in Fig.~\ref{fig:time_dependence_spin_fluctuation}(a) again for $ 0 \leq \omega_{\rm opt} t \leq 224 $. The transitions proceed in two steps if the electric field is applied long enough. With fields weaker than those shown here, the ionicity hardly changes and the system remains neutral. With weak fields shown here, only the first transition appears. With the strongest field shown here, the second transition also appears before the field is turned off. The second transition shows a larger increment of the ionicity and is very similar to the transition shown in the previous subsection, except for the presence of spin fluctuations. 
Then, we here focus on the first transition, where the ionicity increases rather rapidly but slightly to the range between 0.3 and 0.4. As the field is strengthened, this small increase occurs earlier. The field is applied in Fig.~\ref{fig:time_dependence_spin_fluctuation}(b) for $ 0 \leq \omega_{\rm opt} t \leq 60.6$, 65.1, 67.3, 69.6, 71.8 and 74.1 with $ N_{\rm ext} $=270, 290, 300, 310, 320 and 330, respectively. Even when the field is turned off during the small but rapid increase of the ionicity, the ionicity is again almost unchanged after the field is turned off. Because the ionicity remains an intermediate value (below 0.4), the linear behavior is again observed, as shown later.

\begin{figure}
\includegraphics[height=6cm]{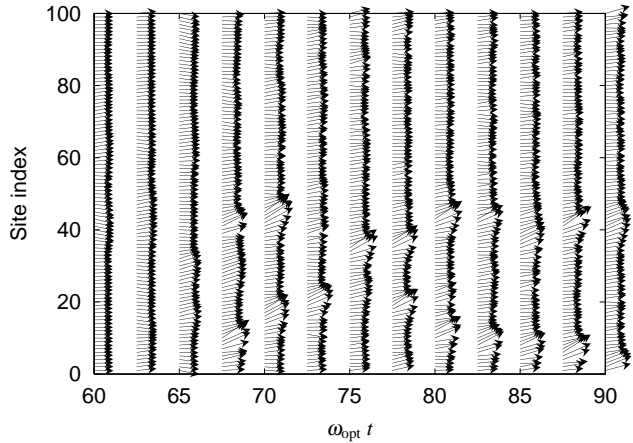}
\caption{Correlation between the staggered magnetization $ m_{st \; l} $ (the vertical component of the bar) and the ionicity $ \rho_l $ (the horizontal component of the bar), as a function of the site index $ l $ and the elapsing time $ t $ multiplied by $ \omega_{\rm opt} $. The electric field with $ eaE_{\rm ext}/\omega_{\rm opt} $=0.6, $ N_{\rm ext} $=300 and of frequency $ \omega_{\rm ext} / \omega_{\rm opt} $=28 is applied (for $ 0 \leq \omega_{\rm opt} t \leq N_{\rm ext} \omega_{\rm opt} T_{\rm ext} $=67.3) to the neutral phase with spin fluctuations at $ T/t $=10$^{-2}$. Two ionic domains appear, but they do not grow after the field is turned off.}
\label{fig:xt_dep_stgmag_short_spin_fluctuation}
\end{figure}

The appearance and evolution of the corresponding partially ionic domain are different from those of the ionic domain without spin fluctuations. The space and time evolution of the ionicity and the staggered {\it magnetization} is shown in Fig.~\ref{fig:xt_dep_stgmag_short_spin_fluctuation}. The horizontal component of each bar represents the local ionicity $ \rho_l $ as before, but the vertical component gives the local staggered magnetization $ m_{st \; l} $ defined as $ m_{st \; l} = (-1)^l ( - m_{l-1} + 2 m_l - m_{l+1} )/4 $, where $ m_l = n_{l,\uparrow} - n_{l,\downarrow} $. The oscillating electric field is applied for $ 0 \leq \omega_{\rm opt} t \leq 67.3 $. Now two domains of somewhat larger ionicity than the background are visible. They are accompanied with spontaneously aligned staggered magnetization. These domains do not grow after the field is turned off, but their positions spatially fluctuate. This fact enables the ionicity to remain an intermediate value if the field is turned off before the transition is completed. Differences between the ionic-to-neutral and neutral-to-ionic transitions are given by whether the metastable domains grow spontaneously (i.e., without electric field) or not. Here also, the staggered lattice displacements remain random, so that ionic domains with different polarizations continue to coexist (not shown). 

\begin{figure}
\includegraphics[height=6cm]{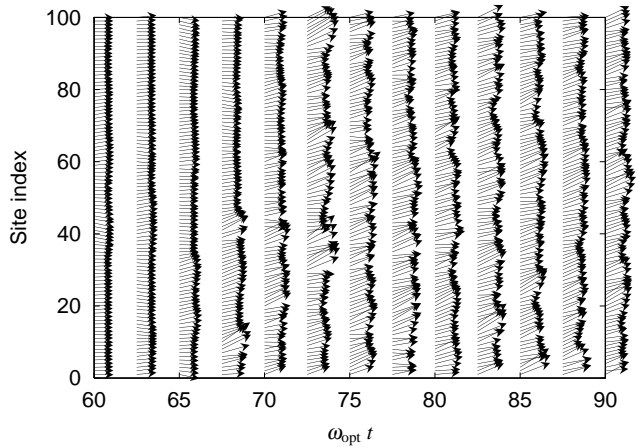}
\caption{Correlation between the staggered magnetization $ m_{st \; l} $ (the vertical component of the bar) and the ionicity $ \rho_l $ (the horizontal component of the bar), as a function of the site index $ l $ and the elapsing time $ t $ multiplied by $ \omega_{\rm opt} $. The electric field with $ eaE_{\rm ext}/\omega_{\rm opt} $=0.6, $ N_{\rm ext} $=330 and of frequency $ \omega_{\rm ext} / \omega_{\rm opt} $=28 is applied (for $ 0 \leq \omega_{\rm opt} t \leq N_{\rm ext} \omega_{\rm opt} T_{\rm ext} $=74.1) to the neutral phase with spin fluctuations at $ T/t $=10$^{-2}$. Ionic domains are produced during the photoirradiation, and they cover the system before the field is turned off.}
\label{fig:xt_dep_stgmag_long_spin_fluctuation}
\end{figure}

The same quantity but with a little bit longer pulse duration is shown in Fig.~\ref{fig:xt_dep_stgmag_long_spin_fluctuation}. Now the oscillating electric field is applied for $ 0 \leq \omega_{\rm opt} t \leq 74.1 $. By the time when the field is turned off, more than two domains of somewhat larger ionicity appear and are overlapped with each other. Their staggered magnetizations are initially out of phase in a narrow region, but they finally become in phase. Although the appearance of the magnetization is an artifact of the approximation, this fact suggests that the symmetry breaking in the ionic phase is not the unique reason for the differences from the ionic-to-neutral transition. In any case, the partially ionic domains finally cover the system so that the ionicity is saturated to be about 0.3. Even if the field continues to be applied, the evolution is almost unchanged in the time span shown here.

\begin{figure}
\includegraphics[height=12cm]{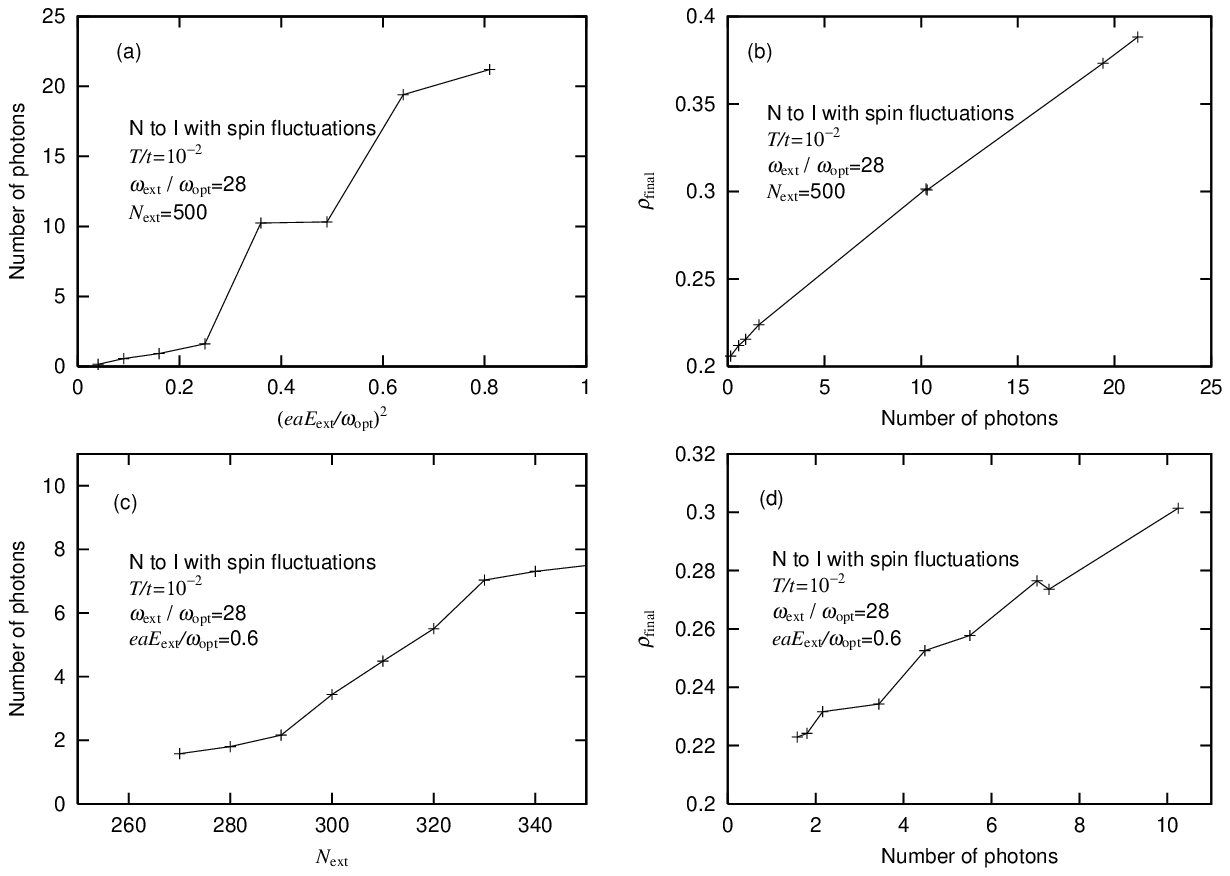}
\caption{(a) Number of absorbed photons as a function of the field strength $ ( eaE_{\rm ext}/\omega_{\rm opt} )^2 $, and (b) the corresponding final ionicity, for $ N_{\rm ext} $=500. (c) Number of absorbed photons as a function of the pulse duration $ N_{\rm ext} $, and (d) the corresponding final ionicity, for $ eaE_{\rm ext}/\omega_{\rm opt} $=0.6. The electric field of frequency $ \omega_{\rm ext} / \omega_{\rm opt} $=28 is applied to the neutral phase with spin fluctuations at $ T/t $=10$^{-2}$.}
\label{fig:pulse_dependence_spin_fluctuation}
\end{figure}

We again calculate the dependence of the number of absorbed photons and that of the ``final'' ionicity (but that before the second transition) on the field strength $ ( eaE_{\rm ext} )^2 $ and the pulse duration $ N_{\rm ext} T_{\rm ext} $, which are shown in Fig.~\ref{fig:pulse_dependence_spin_fluctuation}. The number of absorbed photons is no more a strongly nonlinear function of the strength or of the duration in the range shown here, and it is substantially smaller than the number of sites. Nevertheless, the ``final'' ionicity is again almost linearly related with the number of absorbed photons. No threshold behavior is observed again. Even quantitatively, the linear coefficient is very similar to that calculated without spin fluctuations shown in Figs.~\ref{fig:pulse_dependence_on_resonance} and \ref{fig:pulse_dependence_off_resonance}. Namely, with different frequencies, strengths and durations of the pulse, we obtain $ \rho_{\rm final} $=0.3 and 0.4 roughly when 10 and 20 photons are absorbed in the 100-site chain. From these results, the inclusion or exclusion of spin fluctuations is certainly inessential to produce the linearity.

\subsection{Effect of electron-lattice couplings}\label{subsec:different_lattice_coupling}

To take a hint about how the differences appear between the cooperative ionic-to-neutral transition dynamics and uncooperative neutral-to-ionic one, we have further performed numerical simulations in somewhat unrealistic models. In the dimerized ionic ground state, the lattice is alternately displaced to have long-range order of polarizations, but not in the photoinduced ionic state in the present model. Thus, the spatial correlation of the staggered lattice displacements might be important. In order to enhance the correlation length, we add an elastic coupling between the next-nearest-neighbor lattice displacements and vary its strength. However, the ionicity remains an intermediate value when the field is turned off before the transition is completed. In this case, the polarizations are still disordered in the photoinduced ionic state. 

The dimerized ionic state has a lower symmetry than the neutral state. However, the photoinduced ionic state in the present one-dimensional model system has the same symmetry with the neutral state. Since the cooperativity is generally weaker than higher-dimensional systems, the failure of the long-range ordering is reasonable. As long as the reflectance measurements after intra-chain charge transfer excitations are concerned \cite{okamoto_unpublished,okamoto_unpublished2} no cooperative character is observed, so that the present one-dimensional model system is basically sufficient for comparisons. The symmetry remains high after the photoirradiation, leading to a short correlation length for the staggered lattice displacements, which may be responsible for the weak driving force toward the transition into the ionic phase. Thus, ordering from the disordered state is difficult at least in one-dimensional systems because a part of the supplied energy would always contribute to thermal fluctuations. 

Next, we decoupled electrons from lattice displacements to put the neutral and ionic phases on the same ground with respect to the symmetry. When the photoinduced dynamics is calculated from the neutral phase, the behavior of the ionic domains is similar to that with electron-lattice coupling. That is, the partially ionic domains never spontaneously grow after the field is turned off, in spite of the fact that the staggered magnetizations appear spontaneously. When the photoinduced dynamics is calculated from the ionic phase, the behavior of the ionicity is again similar to that with electron-lattice coupling. That is, the ionicity either reaches a value in the neutral phase or almost returns to the initial value, but it does not take an intermediate value after the field is turned off. Of course, the transition now proceeds much faster because it is not slowed down by lattice displacements. These results imply that the difference in the symmetry between the dimerized ionic and regular neutral states is not a unique reason for the different transition dynamics. Thus, it is suggested that the difference in the origin of the charge gap is relevant: it is the electron-electron interaction in the ionic state, while it is the band structure in the neutral state.

\section{Conclusions}\label{conc}

It is recently and experimentally demonstrated that, in the quasi-one-dimensional mixed-stack organic charge-transfer complex, TTF-CA, the transition from the neutral phase to the ionic phase with three-dimensionally ordered polarizations is induced by intramolecular excitations \cite{collet03}. Such a transition accompanied with symmetry breaking has not so far been observed for intra-chain charge-transfer excitations \cite{okamoto_unpublished,okamoto_unpublished2}. It is probably related to the simultaneously observed, uncooperative character in the latter experiments. To study the neutral-to-ionic transition from the dynamics of charge density coupled with lattice displacements, we again employ the one-dimensional extended Peierls-Hubbard model with alternating potentials. We numerically solve the time-dependent Schr\"odinger equation for the mean-field electronic wave function, simultaneously with the classical equation of motion for the lattice displacements. 

When the neutral phase with almost equidistant molecules is photoexcited, the characters of the transition are qualitatively different from the previously studied ones when the dimerized ionic phase is photoexcited. The final ionicity is almost a linear function of the increment of the total energy and thus of the density of absorbed photons. The energy supplied by photoexcitations is merely consumed to transfer electrons from the donor to acceptor molecules rather independently. This uncooperative character is consistent with the experimental observations for the intra-chain charge-transfer excitations \cite{okamoto_unpublished,okamoto_unpublished2}. The linear coefficient in the relation between the final ionicity and the increment of the total energy is almost independent not only of the strength or duration of the pulse but also of whether spin fluctuations are considered or not at the mean-field level. When the final ionicity is plotted as a function of the amplitude or duration of the pulse, we found that the pulse with frequency at the linear absorption peak is the most effective, which is also a consequence of the uncooperative character. 

The phase transition dynamics is not described by spontaneous growth of a metastable domain. When the electric field is turned off during the transition, the ionicity remains intermediate between those in the neutral and ionic phases. If we allow energy dissipation into electrons in different orbitals or different lattice vibrations from those which are considered in the present model, the electronic state must return to the initial neutral one owing to the uncooperative character. We discussed the origin of the qualitative differences between the ionic-to-neutral and neutral-to-ionic transitions by modifying or removing the electron-lattice coupling. The present results suggest that the difference in the symmetry is not a unique reason but the difference in the origin of the charge gap is important. 

Because the mean-field approximation deals with the Slater determinant, we cannot quantitatively treat the electron correlation by which up- and down-spin electrons repel each other. This constraint increases the total energy, which may effectively give excess energy to disorder the polarizations further. In this sense, the mean-field approximation might be disadvantageous to describe the neutral-to-ionic transition. Although we cannot exclude this possibility, we have shown that spin fluctuations do not alter the linear coefficient in the relation between the final ionicity and the increment of the total energy within the mean-field approximation.

\section*{Acknowledgement}

The author is grateful to S. Koshihara, T. Luty and H. Okamoto for showing their data prior to publication and for enlightening discussions. 
This work was supported by Grants-in-Aid for Scientific Research (C) (No. 15540354), for Scientific Research on Priority Area ``Molecular Conductors'' (No. 15073224), for Creative Scientific Research (No. 15GS0216), and NAREGI Nanoscience Project from the Ministry of Education, Culture, Sports, Science and Technology, Japan.

%------

%-----

\end{document}